\documentclass[fleqn,10pt]{wlscirep}
\usepackage[utf8]{inputenc}
\usepackage[T1]{fontenc}
\usepackage{amsmath}
\usepackage{rotating} 
\usepackage{caption}
\usepackage{xr}

\usepackage{xcite}

\makeatletter
\newcommand*{\addFileDependency}[1]{
  \typeout{(#1)}
  \@addtofilelist{#1}
  \IfFileExists{#1}{}{\typeout{No file #1.}}
}
\makeatother

\title{Towards General Text-guided Image Synthesis for Customized Multimodal Brain MRI Generation}

\author[1,2,+]{Yulin Wang}
\author[2,+]{Honglin Xiong}
\author[2,+]{Kaicong Sun}
\author[2,3]{Shuwei Bai}
\author[2]{Ling Dai}
\author[6]{Zhongxiang Ding}
\author[2]{Jiameng Liu}
\author[2,3,*]{Qian Wang}
\author[1,5,*]{Qian Liu}
\author[2,3,4,*]{Dinggang Shen}
\affil[1]{School of Biomedical Engineering and State Key Laboratory of Digital Medical Engineering, Hainan University, Haikou, 570228, China}
\affil[2]{School of Biomedical Engineering and State Key Laboratory of Advanced Medical Materials and Devices, ShanghaiTech University, Shanghai, 201210, China}
\affil[3]{Shanghai Clinical Research and Trial Center, Shanghai 201210,
China}
\affil[4]{Shanghai United Imaging Intelligence Co. Ltd., Shanghai, 200230, China}
\affil[5]{Key Laboratory of Biomedical Engineering of Hainan Province, One Health Institute, Hainan University, Haikou, 570228, China}
\affil[6]{Department of Radiology, Affiliated Hangzhou First People's Hospital, Xihu University School of Medicine, Hangzhou, 310030, China}

\affil[+]{These authors contributed equally to this work}
\affil[*]{qianwang@shanghaitech.edu.cn, qliu@hainanu.edu.cn, dgshen@shanghaitech.edu.cn}

\begin{abstract}
Multimodal brain magnetic resonance (MR) imaging is indispensable in neuroscience and neurology. However, due to the accessibility of MRI scanners and their lengthy acquisition time, multimodal MR images are not commonly available. Current MR image synthesis approaches are typically trained on independent datasets for specific tasks, leading to suboptimal performance when applied to novel datasets and tasks. Here, we present TUMSyn, a \textbf{T}ext-guided \textbf{U}niversal \textbf{M}R image \textbf{Syn}thesis generalist model, which can flexibly generate brain MR images with demanded imaging metadata from routinely acquired scans guided by text prompts. To ensure TUMSyn's image synthesis precision, versatility, and generalizability, we first construct a brain MR database comprising 31,407 3D images with 7 MRI modalities from 13 centers. We then pre-train an MRI-specific text encoder using contrastive learning to effectively control MR image synthesis based on text prompts. 
Extensive experiments on diverse datasets and physician assessments indicate that TUMSyn can generate clinically meaningful MR images with specified imaging metadata in supervised and zero-shot scenarios. Therefore, TUMSyn can be utilized along with acquired MR scan(s) to facilitate large-scale MRI-based screening and diagnosis of brain diseases.

\end{abstract}

\begin{document}
\captionsetup[figure]{labelfont={bf},labelformat={default},labelsep=period,name={Fig.}}
\flushbottom
\maketitle
%
%
\thispagestyle{empty}

Multimodal and high-resolution brain magnetic resonance imaging (MRI) provides unparalleled opportunities to aid clinical diagnosis, study the intricate human brain structures and functions, and facilitate neurological understanding through its excellent soft-tissue contrast and non-invasive nature. However, the acquisition of multimodal high-resolution magnetic resonance (MR) images is slow due to the imaging mechanism of MRI. 
Furthermore, the scarcity and high costs of MRI scanners further limit the availability of multimodal high-resolution MR images. Hence, to date, multimodal MRI (even critical for assisting disease diagnosis) is often employed only for non-urgent and difficult-to-judge cases in clinics. 

Pioneering works resort to generative models~\cite{sharma2019missing,wang2022deep,zhang2022ptnet3d,liu2023one} and super-resolution (SR)~\cite{zhao2020smore,yurt2022progressively,chen2023deep,feng2024exploring} 
to enrich MR image acquisition in terms of augmented contrasts and improved spatial resolution, respectively. Specifically, medical image synthesis by generative models is employed to impute unavailable MR sequences from the available ones, while SR is adopted to enhance the spatial resolution from the acquired low-resolution (LR) counterpart. However, there still exist noticeable gaps between the performance of generative models and clinical demands, since the existing generative models for MRI synthesis are often trained for specific MRI sequences, 
which are practically inefficient for clinical usage. 
It would be largely beneficial to develop a universal synthesis framework for brain MRI, which can be effectively applied to all the commonly used clinical scenarios, guided by customized imaging parameters. 
Besides, 
the existing generative models mainly focus on modality transfer, ignoring the relevance of spatial resolution. 

The recent advent of generalist models has opened in a new era of generative artificial intelligence (AI)~\cite{brown2020language,alayrac2022flamingo}. These models trained on large-scale datasets can be rapidly adapted to previously unseen tasks and data domains through their mastered general knowledge, and demonstrate remarkable flexibility in handling diverse modalities, such as images and texts. 
When applied to natural image synthesis, several studies have yielded promising results ~\cite{hu2024instruct,tumanyan2023plug}.
Generalized medical artificial intelligence (GMAI) takes advantage of these strengths and applies them to the medical domain~\cite{moor2023foundation}. Several GMAI models are developed for different medical imaging tasks across multiple modalities~\cite{ma2023towards,zhou2023foundation,pai2024foundation,cox2024brainfounder}. 
For example, Brainfounder~\cite{cox2024brainfounder}, trained on a multi-modal MRI dataset from 41,400 participants across three public databases, focuses on two brain abnormality segmentation tasks. Additionally, some models~\cite{lu2024visual,christensen2024vision,zhang2023biomedclip,wu2023towards} adopt contrastive language-image pre-training (CLIP) strategies~\cite{radford2021learning, jia2021scaling} to utilize aligned text and image embeddings to tackle downstream tasks. RadFM~\cite{wu2023towards} aims to tackle a wide spectrum of clinical radiology tasks, including medical visual question answering, radiology report generation, modality recognition, and disease diagnosis, by learning from diverse paired radiological scans and reports. 

Considering the deployment of large-scale networks in clinical practice and the precision required for diagnosis and treatment, some studies~\cite{liu2023clip, hamamci2023generatect} propose generalist models designed for targeted tasks such as image segmentation or disease classification, which are tailored to specific medical imaging modalities. For instance, a CLIP-driven universal model~\cite{liu2023clip} integrates text embedding into CT image segmentation for multiple organs and tumors, which is the first time that clip embedding is used to guide voxel-level medical image analysis task. 
GenerateCT~\cite{hamamci2023generatect} proposes a text-conditional 3D chest CT volume generation framework for data augmentation. However, it is specifically designed for CT images and developed using single-source data. Consequently, it lacks specialized knowledge of MR imaging parameters, and cannot be directly employed for MRI synthesis. 
To the best of our knowledge, there is no existing study for unified brain MRI synthesis, which covers large spectrum of MRI sequences and is guided by ~\textit{\textbf{MR imaging metadata}} as text prompt. 

\begin{figure}[ht]
\centering
\includegraphics[width=\linewidth]{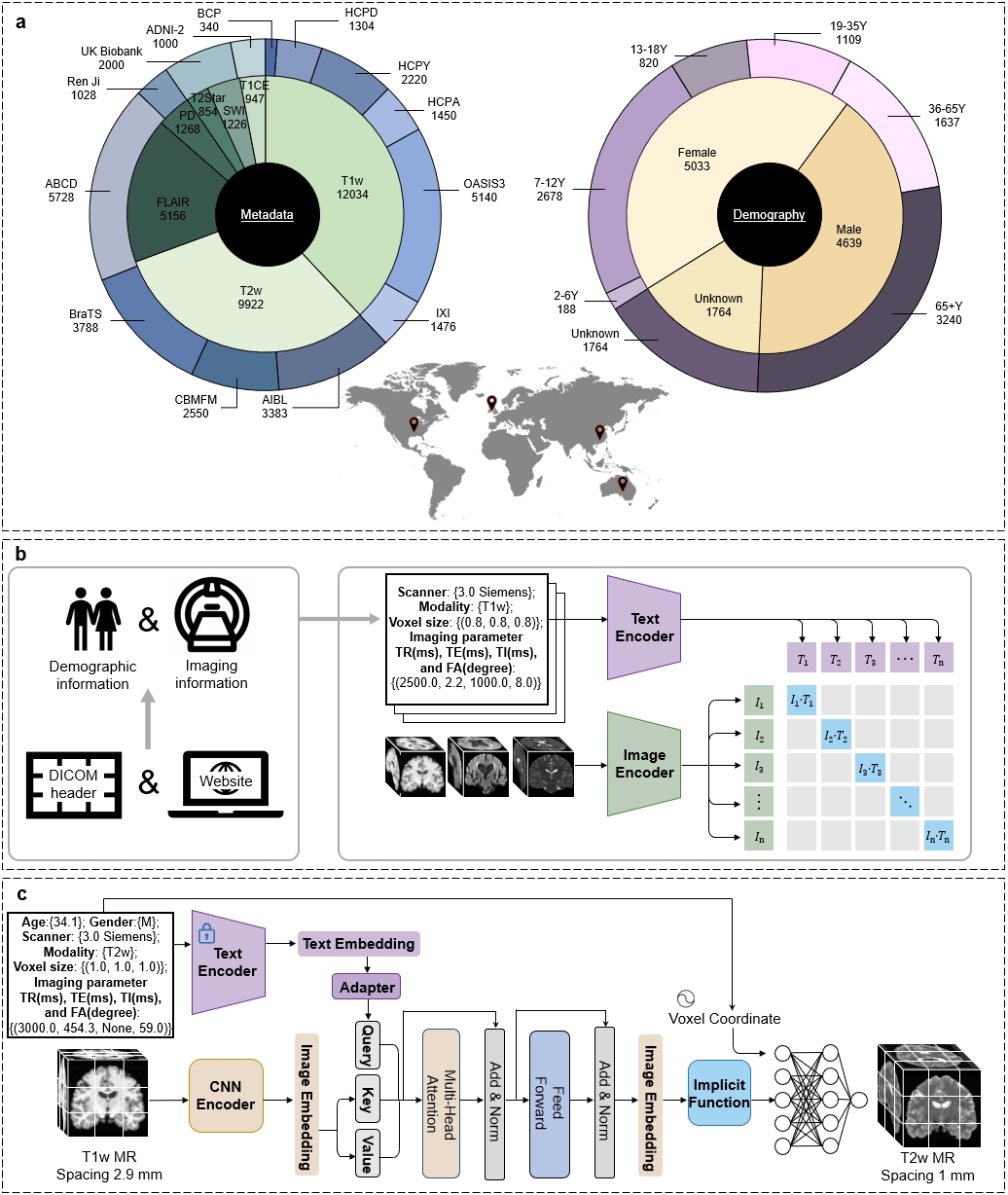}
\caption{Overview of our study. \textbf{a}, Distribution of metadata (dataset and MRI modality) and demographic information (age and gender) along with the amount of images for each category in our database. The geographical representation of data distribution is presented below. 
\textbf{b}, Workflow of pre-training text encoder, showing the construction of text prompts and the use of CLIP for aligning embeddings from text encoder with the ones from image encoder. \textbf{c}, Workflow of training image synthesis model. The pretrained text encoder in Fig.1 b is frozen during the training in this stage. The CNN encoder, dealing with cropped patches instead of the whole volume, is distinct from stage Fig.1 b and is trained as image encoder in this stage. Cross-attention is adopted to integrate text embeddings into image embeddings to control image synthesis. The local implict image function (LIIF) is employed to decode the embeddings and generate images with arbitrary upscaling factors.}
\label{fig:Fig.1}
\end{figure}

In this work, we present the \textbf{T}ext-guided \textbf{U}niversal \textbf{M}R image \textbf{Syn}thesis (TUMSyn) framework, designed to generate customized MRI sequences from routinely-acquired scans based on text prompts (Fig.~\ref{fig:Fig.1}).
TUMSyn is developed using a large database of 31,407 MR image-text pairs from 13 datasets (Table \ref{dataset}). To allow customized synthesis of MRI sequences, 
we pre-train a text encoder to extract embeddings of MR imaging parameters as prompts and align these text embeddings with the corresponding images using the CLIP approach.
We have conducted comprehensive evaluations across a wide range of cross-modal synthesis tasks. TUMSyn consistently surpasses the models trained for specific tasks.
In addition to promising synthesis performance on internal data, evaluation on four external datasets further demonstrates the generalizability of TUMSyn. 
Notably, in zero-shot settings, radiologists' assessments and various evaluation metrics indicate that TUMSyn produces high-fidelity sequences that can meet diverse clinical and research needs, assisting neuro-disease diagnosis and also facilitating brain morphological analysis.  
In summary, guided by text prompts, TUMSyn enables multimodal imaging by effectively generating MRI sequences that are difficult or impossible to acquire in reality, 
providing the potential to significantly augment the efficiency and efficacy of the healthcare system.

\section*{Results}
In this section, we evaluate the synthesis ability of TUMSyn as well as its feasibility to integrate into clinical and research workflows from three perspectives including
1) Tailored MR image generation; 
2) Inherent flexibility and generalizability; 
3) Clinical utility. Details are presented below. 

\subsection*{Text-guided universal framework for brain MRI synthesis}
The overview of our work is shown in Fig.\ref{fig:Fig.1}. The goal of our work is to develop a universal synthesis framework for brain MR images, which can generate desired high-quality MR images according to practical demands provided by text prompts. To ensure the versatility and generalizability of our synthesis model, we collected a large-scale brain MR database with 31,407 3D image-text pairs from 13 datasets located in four continents, including 7 structural MRI modalities, spanning ages from 2 to 100+ years old, and covering a large spectrum of diseases and health conditions (Table \ref{dataset}). 
The text (metadata) includes subject information such as age and gender, as well as key MR imaging parameters, including scanning field strength, scanner type, voxel size, repetition time (TR), echo time (TE), inversion time (TI), and flip angle (FA). The designed English prompt integrates metadata as illustrated in Fig.\ref{fig:Fig.1} b and c. 
To effectively align and fuse image-text pairs, our TUMSyn is built upon a two-stage training strategy. In the first stage, 
we pre-trained a text encoder using contrastive learning to effectively extract textual semantic features which are aligned with the corresponding image features from metadata (Fig.\ref{fig:Fig.1} b).
In Supplementary Fig.\ref{fig:su_fig1} a and b, we evaluate the pre-trained text encoder by showing zero-shot performance on image-to-text retrieval. 
We can see that our text encoder accurately provides highly relevant textual descriptions for the given images, either producing complete prompts or solely imaging modality, 
indicating its ability to understand and master the semantic relationship between paired images and texts. More details are given in section ``Description of experimental setup''. 

Built on the pre-trained text encoder, 
in the second stage, the text encoder is frozen and used to extract prompt features to steer the cross-sequence synthesis. To enable MRI synthesis with desired spatial resolution, the fused text and image features are passed to a decoder, which supports super-resolution for continuous upsampling factor using Local Implicit Image Function (LIIF)~\cite{chen2021learning}. 
In such way, our model, developed on large-scale datasets, can effectively generate target sequences with desired spatial resolution (Fig.\ref{fig:Fig.1} c). 

\begin{figure}[ht]
\centering
\includegraphics[width=\linewidth]{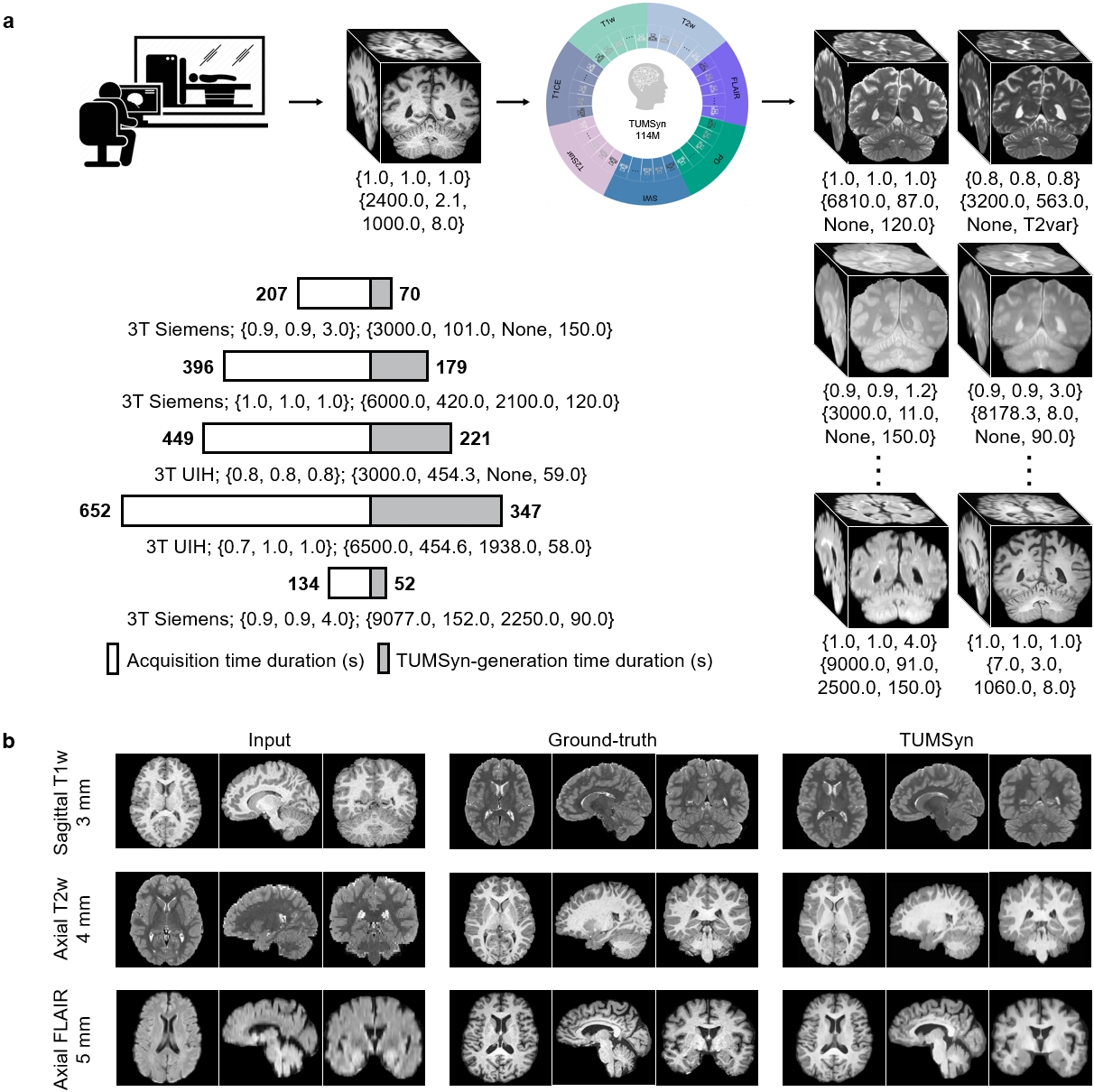}
\caption{Application of TUMSyn in clinical workflow to supplement MRI scanning. \textbf{a}, Integrating TUMSyn into MR imaging workflow. 
TUMSyn, with only 114M parameters, can be easily employed to generate unacquired MRI sequences, governed by text prompts, from seven commonly used MRI sequences in clinics across diverse MRI scanners. Examplar text prompts of target 3D MR images are shown below. The elements in the first curly bracket represent the demanded voxel size of target images, and the elements in the second curly bracket represent the target MR imaging parameters including TR (ms), TE (ms), TI (ms) and FA (degree). 
The tornado diagram in bottom left of Fig.2 a shows execution time of real MRI scans (left) and image synthesis by our TUMSyn (right) for five representative MRI sequences, with corresponding imaging parameters listed below each bar. \textbf{b}, Synthesized images by our TUMSyn from multi-contrast inputs with diverse scanning orientations and spatial resolutions in multiple data centers.} 
\label{fig:Fig.2}
\end{figure}

The abovementioned merits endow TUMSyn with great potential to be integrated into MRI-assisted clinical workflows (Fig.\ref{fig:Fig.2} a). Specifically, once the MRI scanner has acquired any routine MR sequence from a subject, TUMSyn can generate complementary MRI sequences with specified imaging parameters in real time, which can potentially greatly benefit the follow-up diagnosis. 
Besides, compared to real MR scanning, TUMSyn reduces imaging duration by 2- to 4-fold while maintaining clinical equivalence with no additional cost. 
To validate TUMSyn's adaptability to heterogeneous real-world scenarios, we showcased the performance of TUMSyn in synthesizing customized images from various inputs (Fig.\ref{fig:Fig.2} b). In all cases, TUMSyn generates promising images via the guidance of text prompts, underscoring its proficiency in accurate image generation with preserved image details across arbitrary upsampling and cross-modality translation tasks.
It is worth noting that the integration of generative models in MR imaging opens up the opportunity for synthesis-empowered MR scanning and MRI-assisted diagnosis. 

\begin{figure}[ht]
\centering
\includegraphics[width=\linewidth]{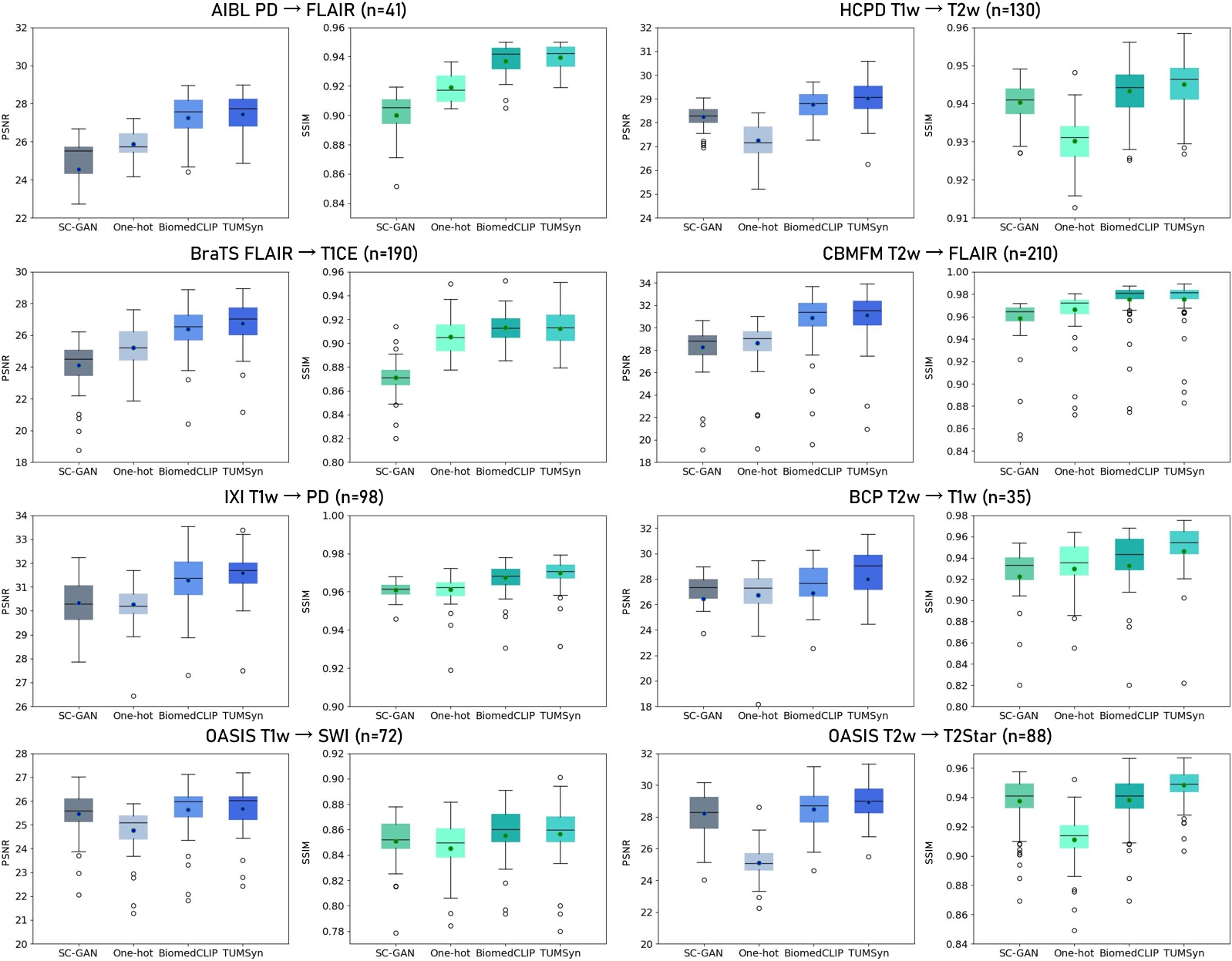}
\caption{Quantitative evaluation of synthesis accuracy and versatility of TUMSyn across eight representative MRI synthesis tasks on internal test sets. Comparison with state-of-the-art synthesis models is conducted in terms of PSNR and SSIM, and shown in barplots with mean, median, upper and lower quartiles.}
\label{fig:Fig.4}
\end{figure}

\begin{figure}[ht]
\centering
\includegraphics[width=\linewidth]{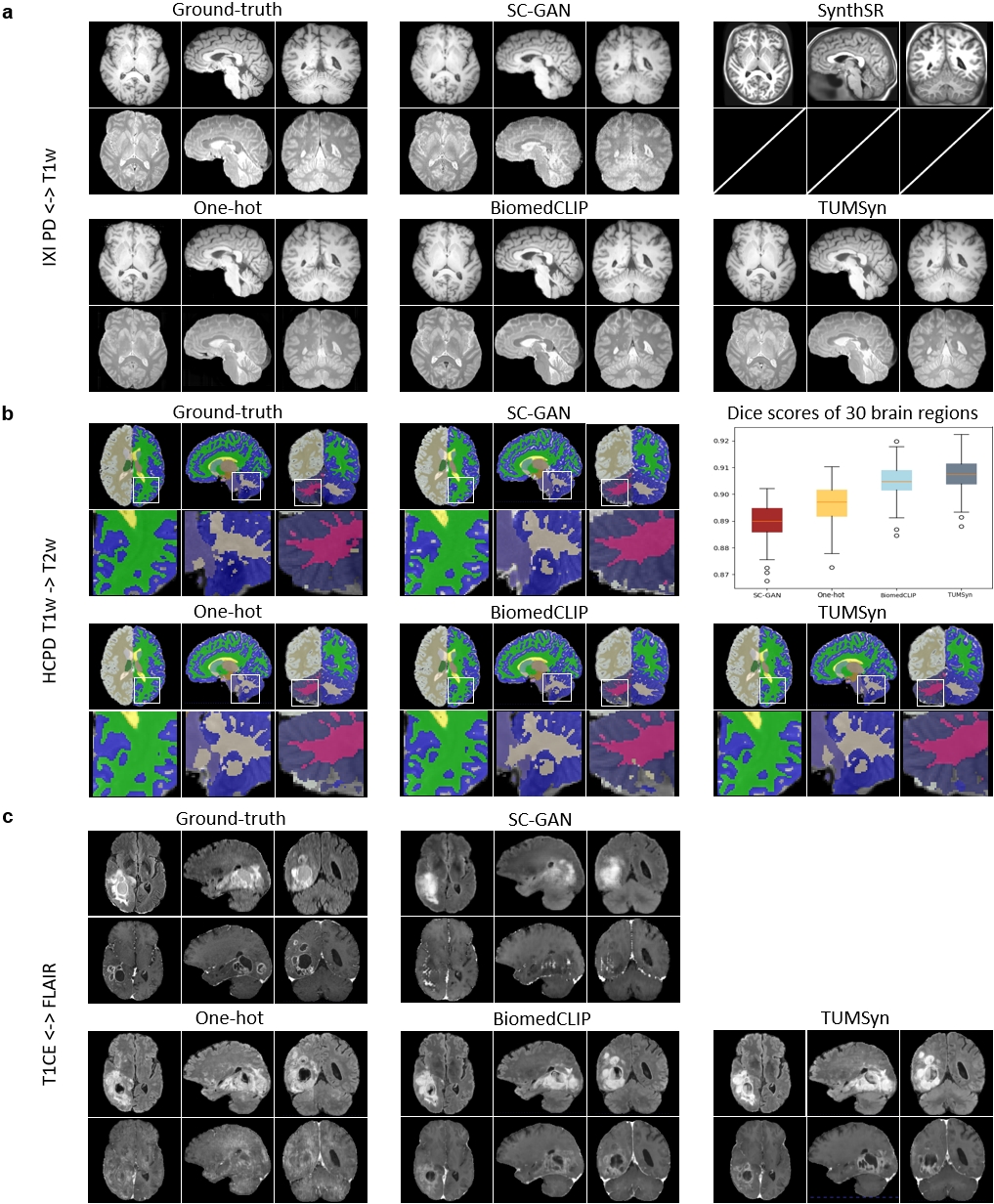}
\caption{Qualitative evaluation of synthesis accuracy and versatility of TUMSyn across multiple MRI synthesis tasks on internal test sets. \textbf{a}, Visual comparison with state-of-the-art methods in synthesis of T1w and FLAIR sequences on IXI dataset. \textbf{b}, Qualitative and quantitative measure for parcellation of 30 brain regions on synthesized T2w images from T1w image pairs by different methods on HCPD dataset. \textbf{c}, Mutual synthesis of FLAIR and T1CE by different methods on BraTs2021 dataset.}
\label{fig:Fig.3}
\end{figure}

\subsection*{TUMSyn in improving synthesis accuracy and versatility}
To study TUMSyn’s synthesis accuracy and ability for diverse tasks using a unified model, we performed a systematic evaluation of TUMSyn on a broad spectrum of synthesis tasks using nine internal datasets.
First, we compared TUMSyn with three competing models, including SC-GAN~\cite{lan2021three}, One-hot model, and BiomedCLIP~\cite{zhang2023biomedclip} model. SC-GAN is a self-attention-based conditional GAN for MR neuroimaging synthesis, and it is a task-specific network trained for independent tasks. 
The One-hot model and BiomedCLIP model, sharing the identical image synthesis model architecture as ours but utilizing one-hot encoding and BiomedCLIP for text embedding, respectively, are employed to evaluate the effectiveness of our pre-trained text encoder on brain MRI synthesis. 
The comparison includes eight synthesis tasks for commonly used MRI modalities on internal test sets (Fig.\ref{fig:Fig.4}). 
The results exhibit that TUMSyn achieves the highest performance for all the tasks. Specifically, TUMSyn significantly outperforms task-specific SC-GAN with both improved PSNR (up to 2.86 dB) and SSIM (up to 0.044). This finding suggests that training on varied data and tasks enables the model to learn generic feature representations, thereby enhancing its overall synthesis ability for different scenarios. 
Besides, compared to the One-hot model, leveraging text prompts can lead to greatly improved synthesis performance, probably because text prompts can precisely instruct imaging parameters of target images.
Moreover, it is shown that, although BiomedCLIP is trained on substantially larger biomedical datasets, our pre-trained text encoder empowers the synthesis model to achieve higher PSNR (up to 1.12 dB) and SSIM (up to 0.010) across eight tasks. This suggests that allowing the text encoder to learn task-related information, namely MR imaging parameters, is critical in enhancing MR image synthesis performance, especially for the downstream tasks that requiring precise prompt features as guidance. 


Visual evaluation is presented in Fig.~\ref{fig:Fig.3}. Besides the aforementioned comparison methods, a general method dedicated to synthesizing T1w images, SynthSR~\cite{iglesias2023synthsr}, is further enrolled into comparison. 
We observe that, since the PD images in the IXI dataset are acquired from two centers using different imaging protocols, the output images of TUMSyn are much more accurate in anatomical details especially compared to SC-GAN, One-hot model, and SynthSR (Fig.~\ref{fig:Fig.3} a). 
These observations once again underscores the importance of utilizing imaging parameters and demographic information as textual prompts to guide accurate image synthesis. 

Besides, 
we further perform whole-brain region parcellation on the HCP development (HCPD) dataset to assess the anatomical structure consistency between synthetic and real-acquired images (Fig.~\ref{fig:Fig.3} b). Totally 30 brain regions are segmented using Synthseg+~\cite{billot2023robust}. We compare the parcellation results by our TUMSyn with those by other methods using visual inspection and Dice score. Both qualitative and quantitative evaluations reveal that the generated images by our TUMSyn enable the most precise parcellation.

In addition, we also evaluate the effectiveness of TUMSyn in handling patients with abnormal brain anatomies. Specifically, we evaluate TUMSyn's performance in synthesizing brain tumor regions using BraTS2021 datasets. 
Compared with other methods on mutual synthesis of FLAIR and T1CE sequences (Fig.~\ref{fig:Fig.3} c), TUMSyn shows superior accuracy in synthesizing contrast enhancements, particularly in delineating boundaries and sizes of the tumor regions, indicating its potential to be applied to patients for assisting disease (i.e., tumor) diagnosis. 

\subsection*{TUMSyn in enabling zero-shot synthesis of multimodal images}

\begin{figure}[ht]
\centering
\includegraphics[width=\linewidth]{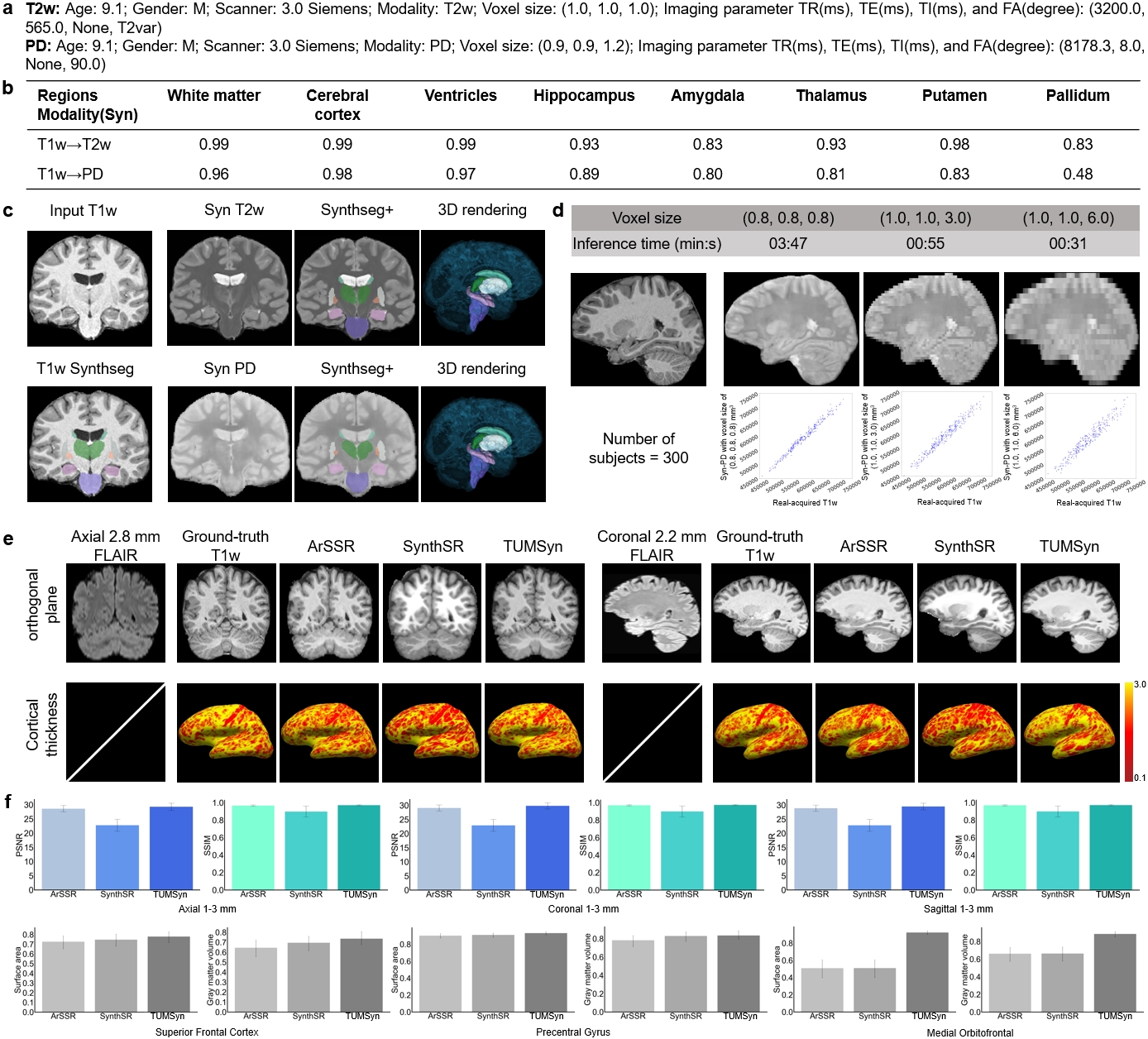}
\caption{Evaluation of TUMSyn on zero-shot synthesis performance. \textbf{a}, Examples of text prompts for synthesizing T2w and PD images from T1w images. \textbf{b}, Pearson correlations for different brain regions parcellated from the generated MR images and the real-acquired T1w images.
\textbf{c}, Visualization of the synthesized T2w and PD images, along with their parcellation results and 3D rendering of 6 brain regions. \textbf{d}, Virtual PD images with various voxel sizes synthesized from T1w image with a voxel size of 1 mm$^{3}$. The table displays the voxel size of synthesized images and the inference times, with the sagittal view of synthesized images showing below. The last row presents the Pearson correlation between the cerebral cortex volume obtained from real-acquired T1w image (x-axis) and the volume obtained from the synthesized PD images (y-axis) at different voxel sizes (Please zoom in to view the details). \textbf{e}, Visualization of real and virtual T1w images by different methods derived from heterogeneous FLAIR images (in orientations and resolutions), along with their cortical thickness. 
\textbf{f}, The first row shows the PSNR and SSIM (mean and standard deviation) of different methods in generating isotropic T1w images from heterogeneous simulated FLAIR images. The second row shows the interclass correlation coefficients (mean and 95\% confidence interval) for surface area and gray matter volume in cortical brain regions of superior frontal cortex, precentral gyrus, and medial orbitofrontal between generated images and real-acquired images 
(Please zoom in to view the details).}
\label{fig:Fig.5}
\end{figure}
Besides synthesis accuracy and versatility, in this section, we systematically evaluate the synthesis generalizability of TUMSyn in handling unseen multi-center data based on zero-shot synthesis on four external datasets.

Firstly, TUMSyn demonstrates its advanced zero-shot synthesis capability on eight cross-sequence image synthesis tasks (Table \ref{tab:table1}). In comparison with other zero-shot methods, TUMSyn consistently outperforms the T1w-synthesis model SynthSR up to 11.10 dB in PSNR and up to 0.135 in SSIM. Similarly, compared to One-hot model and BiomedCLIP model, our TUMSyn achieves a performance gain up to 5.78 dB and 3.22 dB in PSNR, respectively, and 0.069 and 0.028 in SSIM, respectively. 
Particularly, it is worth noting that SC-GAN is trained on each individual external dataset and used as benchmark. We can see that our TUMSyn achieves close zero-shot performance as SC-GAN, and it even surpasses SC-GAN (improvement of 2.51 dB in PSNR and 0.023 in SSIM) for mutual synthesis between T1w and FLAIR sequences on the ADNI-2 dataset. 

Besides promising synthesis performance on MRI sequences that are acquired in practice, TUMSyn also exhibits its potential to supplement MRI sequences that are not acquired in real-world scenarios. We adopt 2,864 scans from the external ABCD dataset, which includes real T2w images while lacks proton density (PD) images. For PD image generation, we use the imaging parameters from the Guy's center in the IXI dataset (Fig.\ref{fig:Fig.5} a). Based on the provided text prompts, TUMSyn generates the T2w and PD images from the corresponding T1w scans. Since there is no ground-truth PD images in the ABCD dataset to directly verify the synthesis quality, we resort to segmentation consistency between the synthesized PD and the associated T1w image pair, measured by Pearson collection (Fig.\ref{fig:Fig.5} b). 
We find that the segmented large regions, including white matter, cerebral cortex, and ventricles, are nearly identical to those obtained from real images. Other regions are also consistent with the real-acquired sequences. For instance, in real-acquired PD sequences, the putamen and pallidum show poor contrast with surrounding gray matter and more blurred boundaries compared to the T2w sequences, and we observe the same trend in our generated images, which is also proved by the decreased segmentation performance for these structures. 
Additionally, we visualize the synthesized PD images along with the corresponding segmentations (Fig.\ref{fig:Fig.5} c), which show plausible structural and contrast characteristics as the PD sequences acquired in the IXI dataset.
Moreover, in Fig.\ref{fig:Fig.5} d, 
we can see that TUMSyn can generate PD sequences at varying voxel sizes with accurate anatomical structures in a much shorter time compared to real scanning. 

Lastly, 
in addition to validating the cross-sequence generation and SR performance of TUMSyn on real input data, we also simulate more diverse input data with various spatial resolutions and scanning orientations, to further demonstrate the model's ability in handling heterogeneous data.
We adopt 300 paired T1w and FLAIR images from the external UK Biobank dataset, and then simulate FLAIR images under various text prompts. Details of the simulation process are available in the Methods section.
Besides SynthSR, TUMSyn is also compared with ArSSR~\cite{wu2022arbitrary}, which is an SR model that can support arbitrary-scale upsampling and particularly trained on this dataset. Notably, across all the three simulated FLAIR images, TUMSyn achieves the best performance in terms of PSNR and SSIM. For example, compared to ArSSR, TUMSyn improves PSNR of images scanned in the coronal direction from 29.08 dB to 29.81 dB. Our method outperforms SynthSR by up to 23.2\% (from 22.91 dB to 29.82 dB) in PSNR (Fig.\ref{fig:Fig.5} f). 
Further analysis on cerebral cortex synthesis, including superior frontal cortex, precentral gyrus, and medial orbitofrontal regions, focusing on cortical surface area (SA) and gray matter volume (GV), exhibit the same trend. The interclass correlation coefficient (ICC) with the 95\% confidence interval (CI) is calculated between the synthesized T1w images and the real-acquired T1w images in these three cortical regions. Morphological analysis shows close matching between TUMSyn-synthesized images and real images with the minimal ICC value of 0.736 (95\% CI 0.666-0.794) and the maximal ICC value of 0.932 (95\% CI 0.911-0.948), demonstrating the ability of TUMSyn in providing superior anatomical structures even in zero-shot inference (Fig.\ref{fig:Fig.5} f). 

\subsection*{Application in detecting neurodegenerative disease-induced hippocampus atrophy}

\begin{figure}[ht]
\centering
\includegraphics[width=\linewidth]{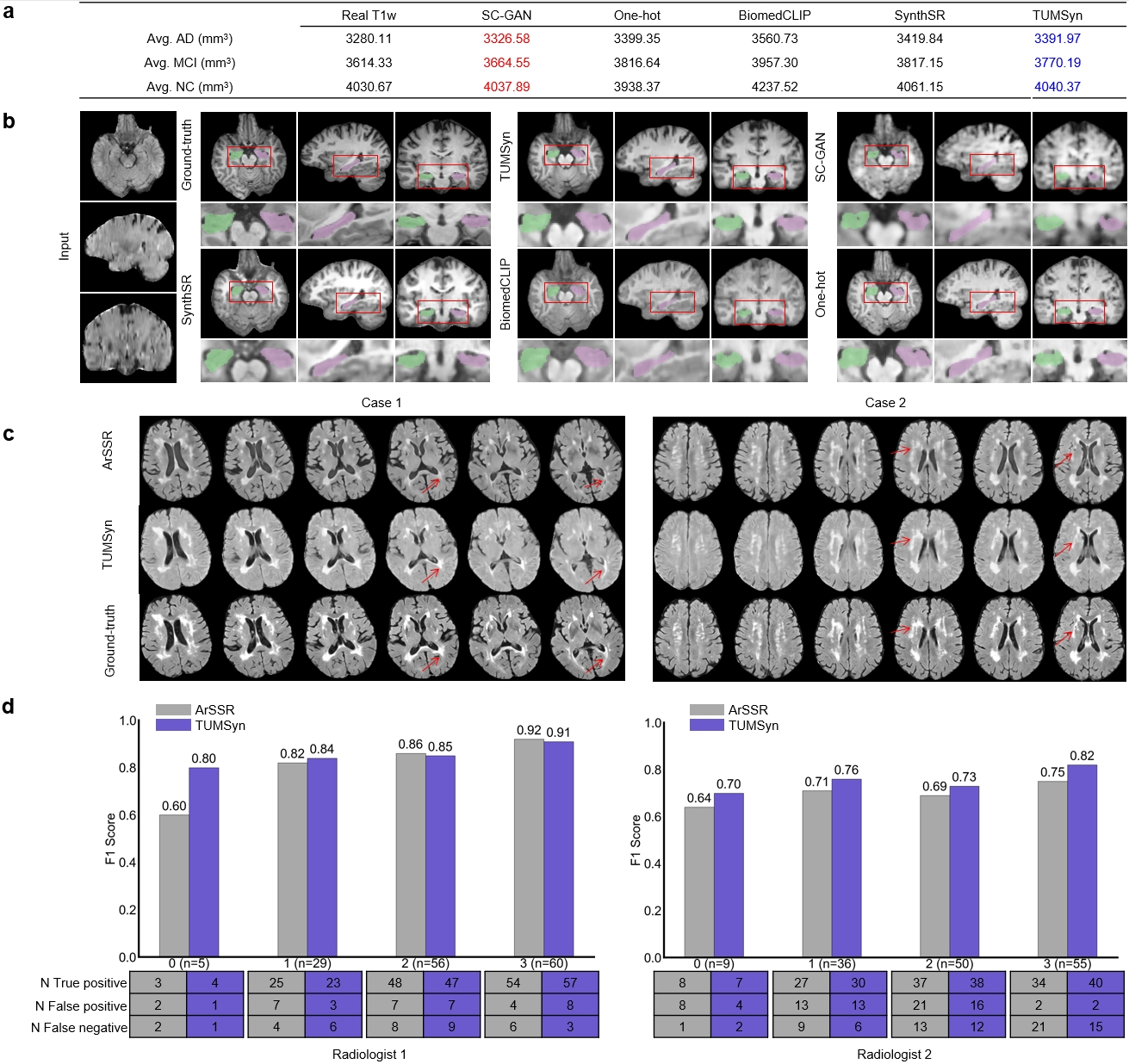}
\caption{Evaluation of TUMSyn-generated images for clinical applications including diagnosis of Alzheimer's disease (AD) and cerebral small vessel diseases (CSVD). \textbf{a}, Average hippocampal volumes estimated from synthesized T1w images by different generative methods and real T1w images for AD, mild cognitive impairment (MCI), and normal controls (NC). \textbf{b}, Synthesized T1w images and the corresponding hippocampus segmentation for an MCI case, with axial FLAIR images as input. \textbf{c}, Evaluation of generated images for CSVD. Two cases of synthesized FLAIR images from T1w images using ArSSR (particularly trained on this dataset) and zero-shot TUMSyn, along with the real-acquired images as ground-truth. For each case, six axial scans generated by each method are presented. Case 1 presents subcortical lesions, and Case 2 shows cortical lesions. The red arrows indicate the lesion areas that physicians use to determine disease grades. \textbf{d}, Fazekas scaling based on synthesized FLAIR images by two radiologists. Compared to real-acquired FLAIR, F1 score for all the Fazekas scales are shown in barplots, with the Fazekas scales and used number of samples displayed below.}
\label{fig:Fig.6}
\end{figure}

Alzheimer's disease (AD) is the most prevalent neurodegenerative disorder worldwide. 
In this experiment, we assessed TUMSyn's capability to generate the shape and volume of the hippocampus, which is a well-established AD biomarker and typically exhibits progressive volume reduction as the disease advances.

We summarize the average segmented hippocampal volumes of AD, mild cognitive impairment (MCI), and normal controls (NC) subjects by different synthesis methods in Fig.\ref{fig:Fig.6} a. The results demonstrate high discriminative power among the three disease stages, closely matching the discrimination exhibited by the real T1w scans. 
Specifically, our approach retains 86\% of the average volume difference between AD and NC (648 versus 751 mm$^3$) and 65\% between MCI and NC (270 versus 416 mm$^3$). 
Furthermore, despite our model did not see FLAIR sequences in ADNI during training, compared with SC-GAN, which is trained on this dataset, TUMSyn achieves at most 3\% drop in accuracy (i.e., SC-GAN loses 1.4\% from 3614.33 mm$^3$ to 3664.56 mm$^3$, while TUMSyn loses 4.3\% from 3614.33 mm$^3$ to 3770.19 mm$^3$ in average hippocampal volume of MCI) for the average segmented volumes across different disease stages. This result indicates that despite the large slice spacing of the input FLAIR data and small volume of the hippocampus, TUMSyn recovers most of the missing high-frequency details and anatomical structures compared to other methods in the zero-shot setting, enabling accurate volume estimation. 
Fig.\ref{fig:Fig.6} b presents an example of synthesized T1w sequences and segmented hippocampus for an MCI patient using different methods. Visually, the images generated by TUMSyn can most accurately delineate the anatomical structures of various brain regions, providing promising contrast between the hippocampus and surrounding areas. In contrast, other methods, because of lacking super-resolution capabilities and flexible generalizability, produce images that are adversely affected by the low-resolution input, and thus over-smoothed synthesis with diminished structural detail. Both qualitative and quantitative observations are inline with the evaluation in Table \ref{tab:table1}.

\subsection*{Application in grading of cerebral small vessel disease}
To validate the potential clinical utility of TUMSyn in identifying white matter hyperintensities (WMH), a type of focal lesion frequently observed in cerebral small vessel diseases (CSVD) and the aging population, we undertake an assessment using 150 subjects in the external dataset from Ren Ji Hospital. Two proficient radiologists conducted Fazekas scale~\cite{fazekas1987mr} grading on (1) real-acquired FLAIR sequences, (2) synthesized FLAIR sequences generated by TUMSyn, and (3) synthesized FLAIR sequences generated by ArSSR (trained on this dataset). The Fazekas scale ranging from 0 to 3 categorizes the severity of the WMH lesion burden from absence to high burden of WMH lesions, respectively.

Fig.\ref{fig:Fig.6} c represents two cases of the axial plane of ground-truth and synthesized FLAIR images by TUMSyn and ArSSR. Surprisingly, in a zero-shot setting, TUMSyn achieves superior lesion synthesis, demonstrating improved lesion morphologies and contrast in both subcortical (Case 1) and cortical (Case 2) regions compared to trained ArSSR. 
According to visual assessments of image quality, radiologists can visually differentiate between real and virtual FLAIR images in most cases. Their discrimination is primarily based in the intensity and texture, rather than the appearance of the lesion regions. To quantitatively evaluate the performance of both methods, we show the F1 scores of the Fazekas scale across four grades between the real FLAIR and the synthesized ones in Fig.\ref{fig:Fig.6} d. The results indicate that the performance of our method surpasses ARSSR in most of cases. Specifically, this superiority is observed in radiologist 1’s evaluation for Fazekas grades 0 and 1, and in radiologist 2’s evaluation for Fazekas grades 0, 1, 2 and 3. We hypothesize that this superior performance may be attributed to the universal learning on large-scale datasets encompassing a diverse range of brain images with abnormalities, including the presence or absence of such WMH patterns. 

\section*{Discussion}

Multimodal brain MR images are beneficial for enhancing diagnostic performance of brain disorders, while are not easily available. In this study, we propose TUMSyn, a unified text-guided brain MR image synthesis framework, to generate text-guided MR images from the acquired MRI sequences, to provide additional MR sequences. 
TUMSyn's foundation rests upon a large scale of training data, encompassing 31,407 3D MR image-text pairs from 13 datasets, covering almost entire lifespan and a large range of structural MR modalities. 
Moreover, TUMSyn achieves accurate and flexible brain MR image generation by pretraining a dedicated medical text encoder harnessing acquisition text/metadata as prompts. 
Altogether, TUMSyn shows~\textit{not only} promising synthesis performance on the internal datasets, \textit{but also} great generalizability on unseen heterogeneous data in the external datasets using a unified model. These attributes position it as a powerful augmentative tool for MRI scanners, bridging the gap between the unavailablity of multimodal MR images in practice and the demand for multimodal MRI assisted disease diagnosis and research. 


Compared to previous medical image synthesis methods, which often rely on task-specific and domain-specific training and have difficulty in dealing with heterogeneous real-world MRI scans, 
our TUMSyn offers a unified solution by utilizing text prompts to effectively steer the synthesis process for various tasks.
Comprehensive evaluation on 9 internal test sets reveals that our TUMSyn achieves consistent superiority over the task-specific models quantitatively and visually. 
Our observations indicate that our unified model TUMSyn with a modest 114M parameters, trained on large-scale datasets for multiple brain MR synthesis tasks, can effectively learn holistic domain knowledge and generic feature representations. Thus it can enhance the synthesis performance for different tasks, revealing great potential to serve as a universal and complementary cross-sequence synthesis tool for brain MR imaging.
Meanwhile, TUMSyn can also provide arbitrary-scale super-resolved MR images by adopting local implicit image function, enhancing the model's utility for diverse imaging scenarios. 


Model generalizability across diverse data domains and tasks is critical for clinical applications. 
We evaluate the zero-shot performance of TUMSyn on four external datasets, and our TUMSyn shows promising synthesis performance, even outperforming the model that is particularly trained on these datasets in some cases.
Besides, by altering textual descriptions of target sequences, TUMSyn can generates high-fidelity sequences with varying imaging protocols, including those not acquired from MRI scanners. These synthesized images ~\textit{not only} fulfill diverse clinical needs, ~\textit{but also} enable detailed volumetric and morphological analyses of crucial brain cortical regions associated with neurological functions.
Moreover, we have validated the effectiveness of TUMSyn on two clinical applications, including diagnoses of Alzheimer's disease and CSVD. 
Both qualitative and quantitative analyses reveal that TUMSyn's performance in predicting hippocampal volume approaches the results of real-acquired images.
A reader study involving two experienced radiologists further validate TUMSyn's clinical utility. For CSVD grading using synthesized scans, the zero-shot TUMSyn achieves comparable or even superior performance across several disease grades, compared to the method particularly trained on the dataset.
TUMSyn's ability to meet various real-world clinical scenarios is particularly valuable, compared to other task-specific methods. Remarkably, TUMSyn achieves this flexibility with negligible costs, presenting the potential to be applied as a supplementary tool in healthcare systems with MRI-related applications. 

We conduct ablation studies to investigate the effectiveness of our brain MRI image-text pre-training model and the use of text prompts. It turns out that our proposed MRI-specific prompts and text encoder consistently outperform both the One-hot model and the BiomedCLIP model, although BiomedCLIP model is generally trained on significantly larger biomedical datasets. This finding indicates that a specifically pre-trained foundational model for dedicated medical tasks, namely brain MRI synthesis in our study, can be more effective and more feasible.  


Besides the clinical applications we have demonstrated, there are other potential applications of TUMSyn for academic research.
For example, TUMSyn could enable the discovery of previously overlooked biomarkers for neurological diseases through rapid MR sequence design and optimization. 
Furthermore, TUMSyn could facilitate large-scale studies, such as investigations on the trajectory of human brain volume changes with age which typically requires substantial sample sizes. TUMSyn could facilitate such large-scale studies by synthesizing diverse and representative brain MRI data, thereby reducing the practical barriers associated with extensive data collection.

While TUMSyn exhibits promising capabilities in text-guided brain MR image synthesis, there are some limitations that are worthy to be noted. The first limitation lies in the architecture design of the synthesis model. 
The rapid progress in deep learning architectures, such as the recent Diffusion Transformer~\cite{peebles2023scalable}, shows the opportunity to further refine TUMSyn, enabling more efficient and precise generation of diverse MR images. Another limitation is that, although the current model successfully provides images with varying contrasts in response to parameter adjustments, TUMSyn may compromise image resolution and contrast for the cases with limited training data or dealing with out-of-distribution data, resulting in visually over-smoothed images. To address this challenge, future efforts may employ larger spectrum of training datasets either by collecting larger-scale MR images or by image augmentation, and further strengthen the alignment between image and text pairs by further regularizing their correlated feature spaces.  

In conclusion, we propose TUMSyn, a general text-guided MR image synthesis framework, to generate customized brain MR images across almost entire age groups and diverse contrasts. 
By exploiting rich datasets and integrating imaging metadata into the image generation process, TUMSyn achieves clinically-meaningful image quality and fulfills diverse real-world application scenarios, 
effectively bridging the gap between the limited MR imaging resources and the clinical demand of multimodal high-resolution MR images. 

\section*{Methods}
\subsection*{Dataset collection and preprocessing}
To establish a large-scale brain MRI dataset, we collected 31,407 3D scans of 12,487 individuals from diverse institutions around the world spanning ages from 2 to 100+ years old across four continents, including Open Access Series of Imaging Studies (OASIS)~\cite{lamontagne2019oasis}, Human Connectome Project (HCP)~\cite{van2013wu}, IXI~\cite{IXI}, Australian Imaging Biomarker and Lifestyle (AIBL)~\cite{ellis2009australian}, Brain Tumor Segmentation (BraTS) Challenge 2021~\cite{baid2021rsna}, Chinese Brain Molecular and Functional Mapping (CBMFM), Adolescent Brain Cognitive Development (ABCD)~\cite{casey2018adolescent}, UK Biobank~\cite{sudlow2015uk}, Ren Ji Hospital in Shanghai, and Alzheimer's Disease Neuroimaging Initiative (ADNI)~\cite{petersen2010alzheimer}, where HCP includes HCP Development (HCPD), HCP Young Adult (HCPY), HCP Aging (HCPA), and Baby Connectome Project (BCP)~\cite{howell2019unc}. The most commonly used MRI sequences in clinics are included such as T1w, T2w, FLAIR, SWI, T2 star, PD, and contrast-enhanced T1w (T1CE) images. Most data used for this study were obtained from publicly available research articles. For in-house datasets CBMFM and Ren Ji Hospital data, the institutional review board approved the retrospective analysis of internal brain MR images. All internal digital data, including MR images and demographic information, were de-identified before computational analysis and model development. Patients were not directly involved or recruited for the study. Informed consent was waived for analyzing MR images retrospectively. The detailed information of each dataset is presented in Table \ref{dataset}, and the scanning parameters along with the corresponding amount of scanns are summarized in Extended Data Fig. \ref{fig:Ex_fig1}.

OASIS is a multimodal dataset generated by the Washington University School of Medicine Knight Alzheimer Disease Research Center (WUSTL Knight ADRC) and its affiliated studies. It has four releases (OASIS-1, OASIS-2, OASIS-3, OASIS-4). In this study, we used the latest release, OASIS-3, which is a longitudinal multimodal neuroimaging, clinical, cognitive, and biomarker dataset for normal aging and Alzheimer's Disease. This dataset contains data from 1,378 participants that were collected across several ongoing projects over the course of 30 years. Participants include 755 cognitively normal adults and 622 individuals at various stages of cognitive decline ranging in age from 42-95 years.

HCP is a five-year project funded by the National Institutes of Health (NIH) and executed by sixteen research institutes or research centers. It was initially launched in July 2009 to build the data on the age group (22-35 years) of subjects and extended to all age groups (BCP, 500 subjects, age 0-5; HCPD, 1,350 subjects, age 5-21; HCPA, 1,200 subjects, age 36-100+) since 2015. The goal of the HCP is to build a ''network map'' that will shed light on the anatomical and functional connectivity within the healthy human brain.

The IXI dataset is led by Imperial College London and comprises nearly 600 MR images collected from normal, healthy subjects. Each subject possesses T1w, T2w, PD, magnetic resonance angiography (MRA), and diffusion-weighted imaging (DWI) sequences with 15 acquisition directions. Data originate from three hospitals using three different MR scanners: a 3T Philips scanner at Hammersmith Hospital, a 1.5T Philips scanner at Guy's Hospital, and a 1.5T GE scanner at the Institute of Psychiatry. Due to the lack of imaging parameters from the Institute of Psychiatry, only data from the first two centers were utilized in this study.

The BraTS challenge is organized by the Medical Image Computing and Computer Assisted Interventions (MICCAI) society for over a decade and has become one of the most popular challenges in the field of medical image processing. This challenge focuses on the evaluation of state-of-the-art methods for segmentation of intrinsically heterogeneous brain glioblastoma sub-regions in multi-parametric magnetic resonance imaging (mpMRI) scans. In this study, we used 947 subjects in the training set of BraTS Challenge 2021, each comprising four MR sequences including T1w, T2w, FLAIR, and T1CE with annotations for background, necrotic tumor core (NCR), peritumoral edema (ED), and enhancing tumor (ET). All data in this dataset have undergone standardized preprocessing, with a uniform spatial resolution of $1.0 \times 1.0 \times 1.0$ mm$^3$ and inter-sequence co-registration. As the provided data is saved in the NIfTI format, the demographic information and imaging parameters are unknown.

AIBL is an ongoing observational cohort study helping researchers unlock new insights into the onset and progression of Alzheimer's disease. This study was launched in 2006, and the dataset was collected at two centers in Perth and Melbourne, Australia. We used a total of 3383 scans, including T1w, T2w, FLAIR, and PD sequences, for training/validation/testing.

The CBMFM dataset was partly collected by the research group in the School of Biomedical Engineering at ShanghaiTech University. This dataset was acquired from four centers, namely Shanghai Fudan University Huashan Hospital, Shanghai Jiao Tong University, the Second Affiliated Hospital of Zhejiang University, and the Shanghai Zhangjiang Brain Intelligence Institute. MR imaging parameters were consistent across all four centers. In this study, we utilized a total of 2550 scans, including T1w, T2w, and FLAIR sequences, for training/validation/testing.

The ABCD study is the largest long-term study of brain development and child health in the United States. The study examines approximately 11,875 youth from 21 sites from age 9 to 10 for approximately ten years into young adulthood. The goal of the ABCD Study is to elucidate the complex interplay between childhood experiences and biological changes, and their impact on brain development. 
The research focuses on a multitude of factors that influence childhood, including engagement with sports, exposure to video games and social media, sleep patterns, and health behaviors such as smoking, and uncovers how these factors influence brain structure and function development, social interactions, behavioral tendencies, academic performance, and overall health outcomes.

UK Biobank is a large-scale biomedical database and research resource containing de-identified genetic, lifestyle, and health information and biological samples from half a million UK participants. It is the most comprehensive and widely-used dataset containing over 40,000 subjects with MR data. In our work, we used 1,000 pairs of T1w and FLAIR sequences for external validation.

Ren Ji Hospital dataset is a longitudinal dataset that includes 598 subjects with no cognitive impairment (NCI) and subcortical vascular mild cognitive impairment (svMCI). This dataset was primarily used to study cerebral small vessel disease and had physician-labeled disease information. In our work, we used 514 pairs of T1w and FLAIR sequences, and we divided them into a training set of 364 and an external validation set of 150 scans, respectively. The training set was also used for the model training of fully- supervised methods for the comparison.

ADNI is a longitudinal multicenter study designed to develop clinical, imaging, genetic, and biochemical biomarkers for early detection and tracking of Alzheimer's disease (AD). This study has recruited participants at 57 sites in the United States and Canada, ranging in age from 55 to 90 years. The initial five-year study (ADNI-1) began in 2004 and was extended by two years in 2009 by a Grand Opportunities grant (ADNI-GO), and in 2011 and 2016 by further competitive renewals of the ADNI-1 grant (ADNI-2 and ADNI-3, respectively). In our work, we used paired T1w and FLAIR images from a total of 150 subjects with normal cognitive aging, mild cognitive impairment (MCI), and early Alzheimer's disease in ADNI-2, each disease stage has 50 subjects respectively.

Among these datasets, OASIS, HCP, IXI, BCP, AIBL, BraTS2021, and CBMFM were used for training and internal validation, and the ABCD, UK Biobank, Ren Ji, and ADNI were used for external evaluation. 

To encourage the model to accommodate practical scenarios from multi-center data, 
the preprocessing step is reasonably simplified to a minimal extent. First, we performed co-registration for each subject and then stripped skulls for each scan. No other operations were conducted. To utilize text prompts, we collected demographic information and imaging parameters including field strength, scanner model, voxel size, TR, TE, TI, and FA of each scan from DICOM header files and official websites of datasets. These imaging parameters control the contrast and resolution of the image during the real MR scanning process.

\subsection*{Model architecture and training}
The model framework is demonstrated in Fig.~\ref{fig:Fig.1} b and c, which contains two training stages: the language-image pre-training (shown in b) and the training of the image synthesis network (shown in c). More details are given below.

\subsubsection*{Image-text pre-training}
The objective of the image-text pre-training model is to build cross-modal correspondences between visual content and linguistic semantics, which enables text-guided image synthesis for the follow-up image synthesis network. 
In this study, to align our text prompts and 3D volumes, we develop a domain-specific Brain MRI Language-Image Pre-training (BMLIP) model, including an image encoder and a text encoder, that can establish relationships between 3D brain MRI scans and their corresponding imaging parameters, namely scanning field strength, MR scanner model, voxel size, TR, TE, TI, and FA.
The text encoder is capable of holding a text length of 90 tokens and is specific to characterizing brain MR imaging parameters. 
Due to the abundance of numerical and sequential information within the text prompts, we adopt the pre-trained Byte-Pair Encoding (BPE) tokenizer used in the original CLIP paper~\cite{radford2021learning}. Compared to other tokenizers, the BPE tokenizer has superior adaptability across a wider range of datasets. <startoftext> and <endoftext> tokens are added at the beginning and the end of each text prompt to distinguish between different text prompts. Subsequently, the numerical sequences are passed into a text encoder to obtain text representations. Due to the richer content in our text prompt compared to the ones used in the original CLIP that only contains category information, we set the encoded length to 90 tokens.
Inspired by the original CLIP, the text encoder is a Transformer with modified architecture~\cite{radford2019language}. 
As the image encoder, we utilize ViT-B/16 architecture as the backbone, where B represents the base model size consisting of 12 transformer layers and 12 attention heads with an embedding dimension of 768 and a hidden dimension of 3,072, and 16 denotes the token size of $16\times16\times16$. Considering the critical importance of inter-slice structural information, we modify the network architecture, which is designed for 2D natural images, to accommodate 3D MR images. To manage the computational demands of 3D image computing, we downscale the entire input volume to $\frac{1}{8}$ of its original size (by half in each dimension) and further crop out the background to the size of 96$\times$96$\times$96 (Extended Data Fig.~\ref{fig:Ex_fig2} a). 

BMLIP is trained using an equal-weighted combination of image-text contrastive loss for the text and image encoders. 
The loss is designed to minimize the cosine similarity between paired samples and maximize the cosine similarity between non-paired samples. The contrastive loss is formulated as:
\begin{equation}
\mathcal{L} = \frac{1}{2N} \sum_{i=1}^N \left( \mathcal{L}_{\text{image}}(i) + \mathcal{L}_{\text{text}}(i) \right)
\end{equation}
where $\mathcal{L}_{\text{image}}(i)$ and $\mathcal{L}_{\text{text}}(i)$ represent the similarity loss for image encoder and text encoder, respectively, and each is defined as 
\begin{equation}
\mathcal{L}_{\text{image}}(i) = -\log \frac{\exp\left(\cos(\mathbf{v}_i, \mathbf{t}_i) / \tau\right)}{\sum_{j=1}^N \exp(\cos(\mathbf{v}_i, \mathbf{t}_j) / \tau)},
\end{equation}
\begin{equation}
\mathcal{L}_{\text{text}}(i) = -\log \frac{\exp(\cos(\mathbf{t}_i, \mathbf{v}_i) / \tau)}{\sum_{j=1}^N \exp(\cos(\mathbf{t}_i, \mathbf{v}_j) / \tau)},
\end{equation}
where $N$ represents the batch size, $\mathbf{v}_i$ and $\mathbf{t}_i$ denote the encoded image and the text features of the $i$-th sample, respectively. The $\cos(\cdot, \cdot)$ operator calculates the cosine similarity between the two features, and $\tau$ is a constant that controls the concentration of the distribution.
The training process consists of 100 epochs, with a warm-up learning rate (from 0 to 5 $\times$ 10$^5$) for the first 20 epochs, and a gradual decrease in learning rate from 5 $\times$ 10$^5$ to 0 for the rest epochs. We use a mini-batch size of 27, representing the category number of imaging parameters in our datasets. Including all classes in each mini-batch ensures sufficient negative samples for effective contrastive learning. 
The model is optimized by the Adam optimizer with $\beta_1$ = 0.5, $\beta_2$ = 0.999. The model variant of the last epoch is saved and employed for downstream tasks. All the training and inference of our model
use a single NVIDIA A100 GPU equipped with 80 GB RAM. 
The total training time is approximately 40 hours.

\subsubsection*{Image synthesis network}
The image synthesis network is designed to generate target MR images from the available images steered by the text prompts. This network comprises four key components including an image encoder, a text encoder, a multi-modality cross-attention module, and an implicit decoder (Fig.~\ref{fig:Fig.1} c). Concerning memory limitation, we 
cut volume into image patches with size of $64\times64\times64$ voxels as input of the image encoder. 
Initially, the image patches and text prompts are encoded into features through their respective encoders. It should be noted that the text encoder is pretrained and frozen in this stage, while the image encoder is trained on a larger backbone. 
Specifically, the image encoder (Extended Data Fig.~\ref{fig:Ex_fig2} b) is built on the24-layer ResNet architecture, omitting downsampling operation to preserve the original resolution of the anatomical structure of brain regions. The image encoder maintains a hidden dimension of 256 in its resblocks, and ultimately projects to a 768-dimensional feature space via a convolution layer. The text features are fed into a text adapter, which consists of a fully connected layer and a rectified linear unit (ReLU) that adapts the features from the frozen text encoder to the image synthesis task. 
The extracted image and text features are then passed to a cross-attention module, which follows the standard multi-head self-attention architecture. In particularly, the text embeddings serve as the Query and image embeddings serve as the Key and Value. The cross-attention module integrates embeddings of both text and image modalities and generates the target image features controlled by the text prompts.

The image decoder aims to achieve arbitrary-scale upsampling (Extended Data Fig.~\ref{fig:Ex_fig2} c). Here, we adopted the LIIF which represents images as continuous functions rather than discrete pixels. Specifically, the image decoder takes the LR image feature maps from cross-attention module as a grid of feature vectors. Each feature vector $F_{x}$ corresponds to a voxel at the coordinate $\textit{X}$ in the original image. For any given coordinate in the continuous image space, LIIF identifies the nearest feature vectors from the grid and performs interpolation according to the upscaling factor to generate a location-aware feature vector. This interpolated feature vector, along with the corresponding target coordinates, is then fed into LIIF decoder to predict the pixel value at that specific location.  The LIIF decoder is constructed with four linear layers, with embedding dimensions of 3072, 3072, 768, and 256, respectively. 
The output of the decoder is ultimately a resolution-enhanced brain MR image guided by prompt-specified imaging sequence.

To train the synthesis network, we use the mean absolute error (MAE) loss to penalize the difference between the synthetic images and the real images. The network is trained for 300 epochs, and the mini-batch size is set as 16. Adam is used as the optimizer with $\beta_1 = 0.5$, $\beta_2 = 0.999$, and the initial learning rate is $1 \times 10^{-4}$ for the first 100 epochs and gradually decayed with a step size of 50 by using the ``MultiStepLR'' module in Pytorch. The training takes about two weeks on an NVIDIA A100 GPU.  

\subsection*{Description of experimental setup}
To evaluate our TUMSyn, we set up the experiments from different perspectives including synthesis accuracy, versatility, generalizability, and clinical utility. The details are given below.
\subsubsection*{Evaluation of synthesis accuracy and versatility}
Regarding synthesis accuracy and versatility, we have evaluated our TUMSyn on nine internal test sets. For comparison methods, we employ SC-GAN, BiomedCLIP, and the One-hot model. Specifically, for SC-GAN, we use the original code provided by the authors of ProvoGAN~\cite{yurt2022progressively}. The batch size is set as 2 and the training epoch is 120. These hyperparameters are selected based on the performance on validation sets. The One-hot model and BiomedCLIP model share the identical synthesis model architecture as TUMSyn. Although one particular modality (e.g., T1w or T2w) from the same center may have different imaging parameters in real-world scenarios, assigning a unique numerical label to each imaging parameter setup is impractical. Therefore, we assign a single numerical label to each modality for the same center during the training of the One-hot model.
The BiomedCLIP model uses BiomedCLIP, which is a language-image foundation model pre-trained on 15 million biomedical image-text pairs, to generate text embedding to guide image transfer. 
For all the T1w synthesis tasks, we additionally involve SynthSR into comparison, which is designed to transfer multiple MR modalities to T1w images and trained using a large amount of synthesized images. Since it has been integrated into Freesurfer~\cite{fischl2012freesurfer}, we did not perform its additional training on our datasets. 

\subsubsection*{Evaluation of synthesis generalizability}
To evaluate the generalizability of TUMSyn, we have conducted zero-shot inference on four external datasets, namely ABCD, ADNI-2, UK Biobank, and Ren Ji. We first synthesize MR image sequences whose ground-truth are available, and then synthesize the image sequences whose ground-truth are unavailable. 
For the former cases, we quantitatively evaluate the synthesized images in terms of PSNR and SSIM as shown in Table~\ref{tab:table1}. For the latter ones, we carry out parcellation of eight brain regions using the segmentation network Synthseg+, which is an AI-based segmentation tool that supports robust analysis of heterogeneous clinical datasets with arbitrary MR contrasts and resolutions. 
Furthermore, we analyze the performance for heterogeneous data with different spatial resolutions and scanning orientations. Heterogeneous FLAIR sequences are simulated with distinct inter-slice spacing by applying randomly selected downsampling factors in the range of [1,3] along the axial, sagittal, and coronal axes, followed by Gaussian filtering with the full-width at half maximum (FWHM) of [1, 3] in UK Biobank dataset. 
Besides SynthSR, the super-resolution model ArSSR, which supports arbitrary upscaling for 3D MR images, is additionally integrated into comparison. 
The original code of ArSSR is used and ArSRR is particularly trained on each of the external datasets to provide performance benchmark. 
The training, validation, and testing sets comprise 1862, 138, and 860 samples for the ABCD dataset, and also 650, 50, and 300 samples for the UK Biobank dataset. The training batch size is 2 and the training epoch is 120 for both datasets. 
For analysis of brain cortical surface generation, we utilize Freesurfer, a widely adopted software package for analysis and visualization of the structural and functional neuroimaging data. Besides, all the subcortical region segmentation is performed using Synthseg+.

\subsubsection*{Evaluation on two clinical applications}
To assess clinical impact of our TUMSyn, we applied TUMSyn for AD and CSVD diagnosis using ADNI-2 and Ren Ji datasets, respectively. 
For AD diagnosis, 
the ADNI-2 dataset is divided into 320, 30, and 150 for training, validation, and testing, respectively. The testing set comprises 50 AD patients, 50 MCI patients, and 50 NC. Each subject has a high-resolution T1w scan (1 mm slice spacing) and an axial FLAIR scan (5 mm slice spacing). We employ Synthseg+ to segment the hippocampus of the synthesized T1w images from the FLAIR input, and compare with the ground-truth segmention from real T1w images.
For the CSVD experiment, the Ren Ji dataset is used, which contains 334, 30, and 150 scans for training, validation, and testing, respectively. 
In both experiments, all the hyperparameters and training configurations are consistent with the aforementioned setups. 

\subsubsection*{Evaluation of pretrained text encoder through image-to-text and image-to-modality retrieval}\label{subsec:retrieval}
To evaluate our text encoder's understanding of the semantic relationship between text and image, we perform image-to-text retrieval following the commonly used evaluation pipeline. Specifically, 
we use our pretrained image and text encoders in the first stage to extract image and text embeddings, respectively. Given an image embedding, we calculate its cosine similarities with 10 randomly selected text prompt embeddings on the test set. Besides, we also perform image-to-modality retrieval, which retrieves only the imaging modality but not the text prompt given the image input. We use all the seven MRI modalities as the candidates of retrieving modalities. Since most of the CLIP models are only applicable to 2D images, such as the original CLIP model and BiomedCLIP, and none of them are pretrained using MR imaging parameters, we do not consider other CLIP models into comparison here. 

\subsection*{Reader study protocol and Fazekas scale}
Reading was performed to evaluate the effectiveness of synthesized FLAIR images for CSVD diagnosis, taking real-acquired FLAIR as the reference. 
Two neuroradiologists, one with 31 years of experience (Z.Ding) and another with 12 years of experience (S.Bai), have analyzed the white matter regions of FLAIR images for 150 patients. Real-acquired FLAIR images and synthesized FLAIR images by two methods are read once by both radiologists for all testing patients, for a total of 450 reads. Two radiologists are instructed to grade the lesion levels on a 4-point Fazekas scale ranging from 0 (no lesion) to 3 (large white matter lesion) and to report the overall image quality of the synthesized images. Scaling is established for each image independently without access to any complementary information. Scaling by the more experienced radiologist based on real-acquired FLAIR images is used as reference.  

For evaluation of white matter lesions, the Fazekas scaling system is employed to assess the severity of periventricular white matter (PVWM) and deep white matter (DWM) abnormalities. The original Fazekas scaling criteria for PVWM and DWM are as follows: ~\textit{Periventricular White Matter (PVWM)}: 0: No lesion; 1: Caps or pencil-thin lining; 2: Smooth halo; 3: Irregular periventricular hyperintensity extending into the deep white matter. ~\textit{Deep White Matter (DWM)}: 0: No lesion; 1: Small punctate lesion; 2: Beginning confluence of lesions; 3: Large confluent lesions. For clinical convenience, a modified Fazekas scaling system is often utilized, where lesions in both PVWM and DWM are collectively assessed without differentiation. In this study, we adopt this modified Fazekas scaling system.

\subsection*{Evaluation metrics}
The Peak Signal-to-Noise Ratio (PSNR) and Structural Similarity Index (SSIM) are used to evaluate the performance of image synthesis. The former (PSNR) is utilized to measure pixelwise error between the synthesized and real-acquired images. 
Higher values indicate better similarity between the synthesized image and the authentic image. The latter (SSIM), ranging from 0 to 1, 
relies on the similarity between synthesized and real-acquired images in terms of luminance, contrast, and structure. 
For SynthSR that automatically synthesizes skull during inference, we strip the skulls and then calculate the PSNR and SSIM. For our method and other comparison methods, we consider the 99.5 percentile of the maximum and the minimum values of the original images as the maximum and minimum values, respectively, and then normalize the image intensities to [0, 1] based on min-max normalization.
The F1 score is used to evaluate the performance of the synthesized images based on disease grading. Ranging from 0 to 1, the F1 score is calculated from the harmonic mean of precision and recall score. Furthermore, to determine the CI, predictions are bootstrapped 100 times, each with 70\% of the data used in the calculation.

\appendix
\captionsetup[figure]{labelfont={bf},labelformat={default},labelsep=period,name={Extended Data Fig.}}
\setcounter{figure}{0}
\begin{sidewaysfigure}[ht]
\centering
\includegraphics[width=\linewidth]{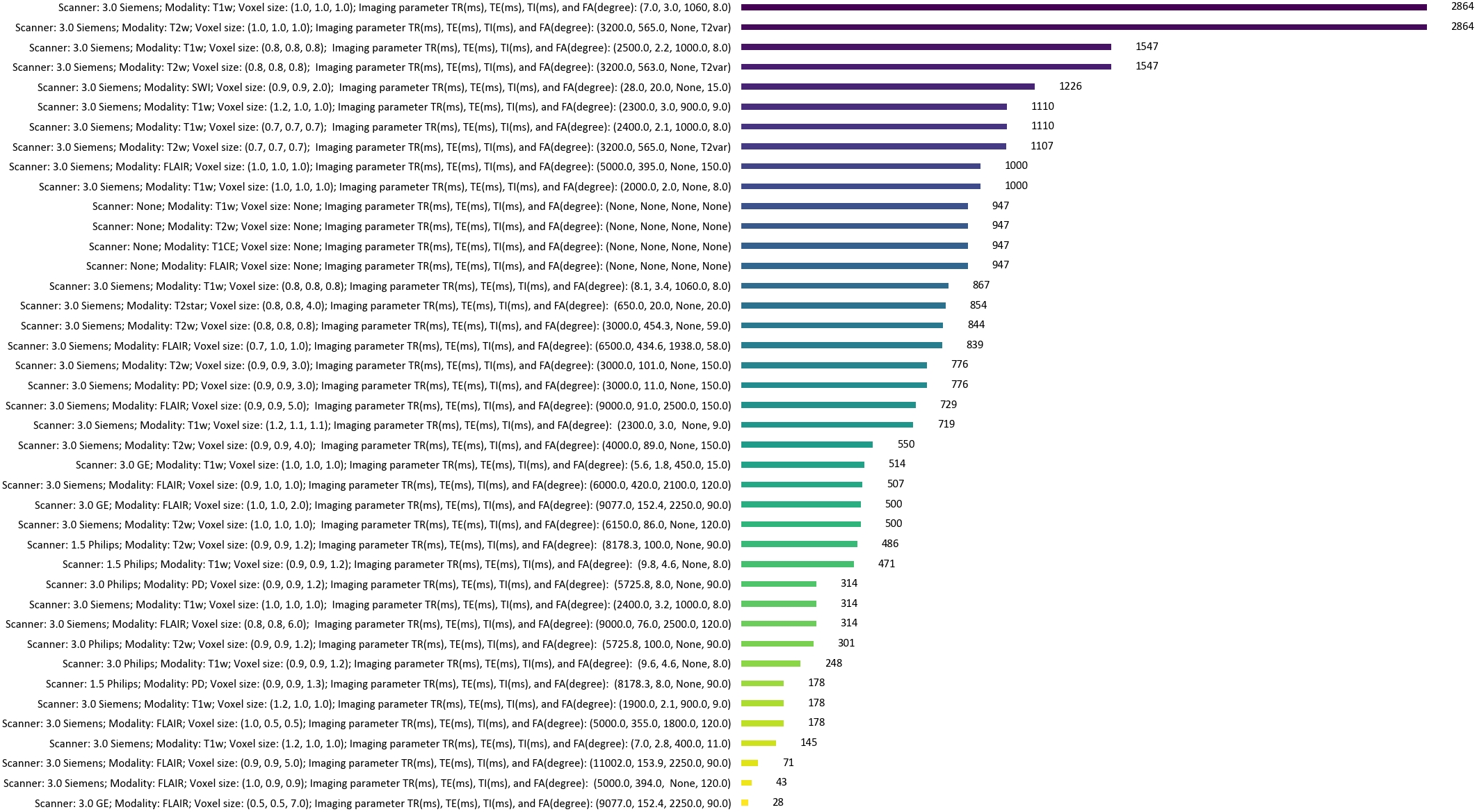}
\caption{Distribution of the imaging parameters across our entire database.}
\label{fig:Ex_fig1}
\end{sidewaysfigure}

\begin{figure}[ht]
\centering
\includegraphics[width=\linewidth]{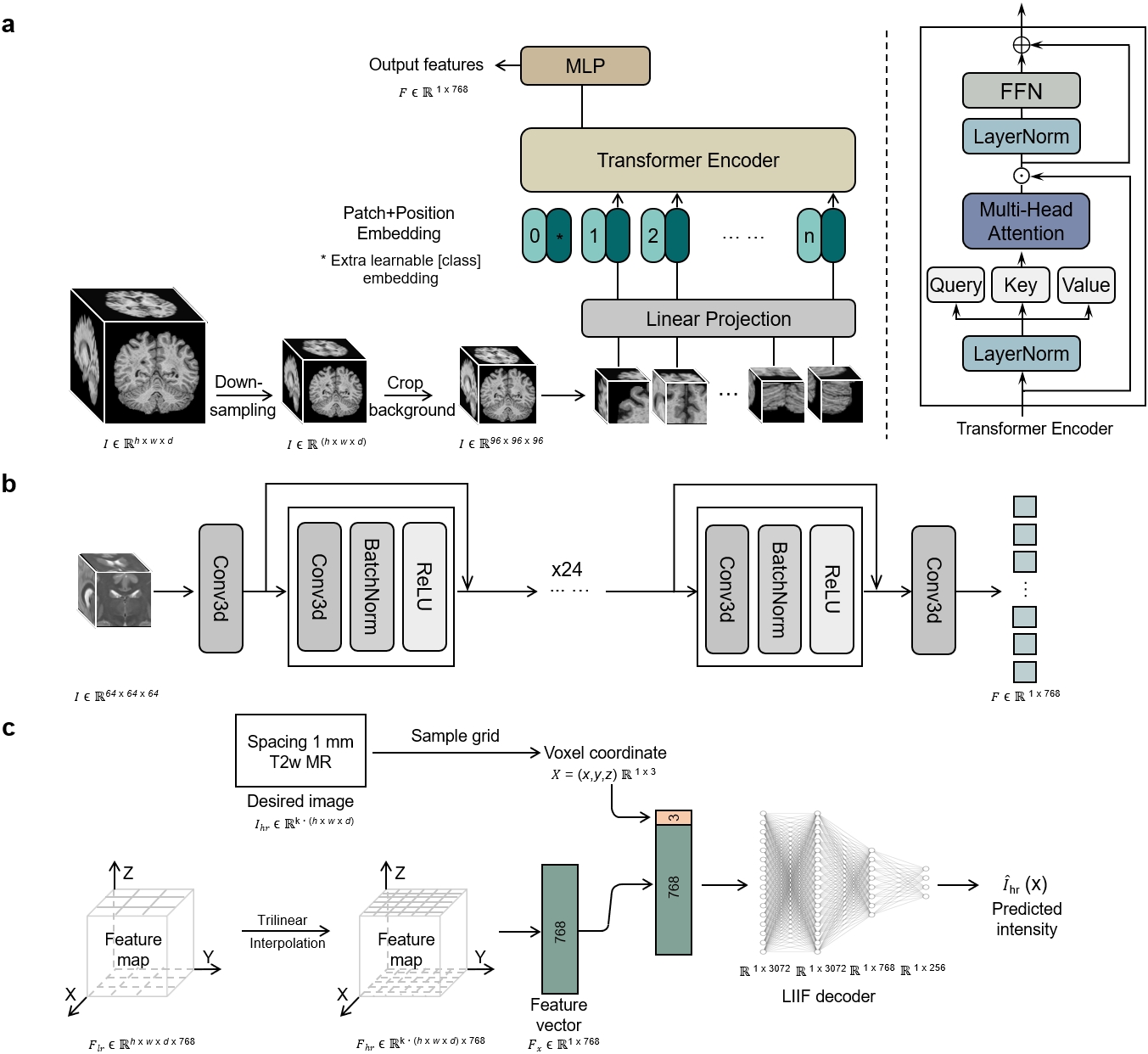}
\caption{Detailed illustration of our model architecture. \textbf{a}, Architecture of the image encoder in BMLIP, which is used for pre-training text encoder. \textbf{b}, Architecture of the CNN encoder in the image synthesis model, which is built on a 24-layer ResNet. \textbf{c}, Architecture of the LIIF-based image decoder in the image synthesis model.}
\label{fig:Ex_fig2}
\end{figure}

\setcounter{figure}{0}
\captionsetup[figure]{labelfont={bf},labelformat={default},labelsep=period,name={Supplementary Fig.}}

\begin{figure}[ht]
\centering
\includegraphics[width=\linewidth]{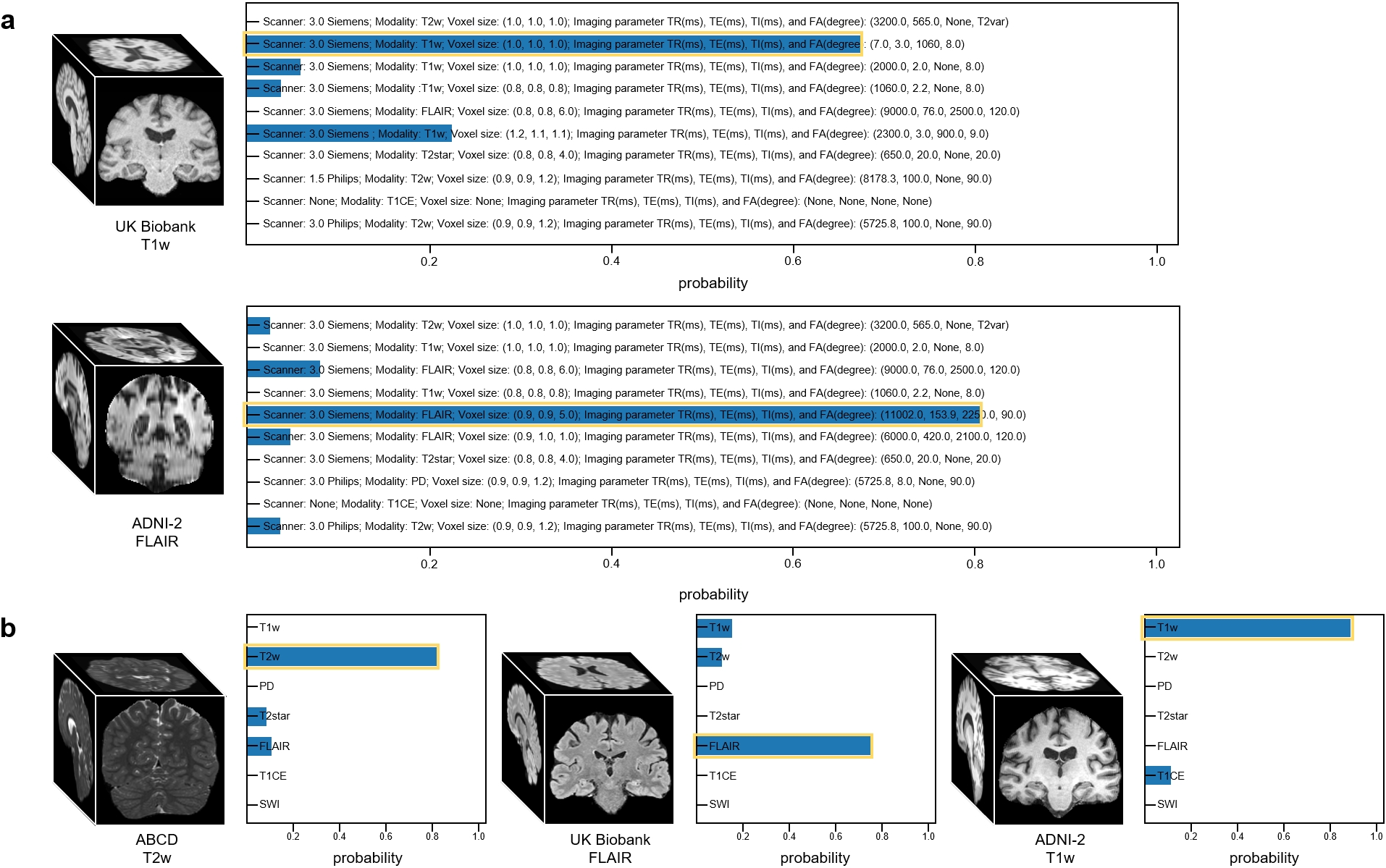}
\caption{Zero-shot performance of our text encoder on (\textbf{a}) image-to-text and (\textbf{b}) image-to-modality retrieval. Given an image as input, the text with highest cosine similarity match with the image embedding is retrieved. For image-to-text retrieval, 10 randomly selected complete text prompts are used and for image-to-modality retrieval, 7 image modalities are used as text prompts. 
We demonstrate several examples for both cases. For all the examples, we denote the retrieval probability using blue bar. 
Gold box indicates the ground-truth text prompt for the given image.}
\label{fig:su_fig1}
\end{figure}

\bibliography{sample}


\section*{Code availability}
The code used in the current study to develop the algorithm is available at https://github.com/Wangyulin-user/TUMSyn.

\section*{Acknowledgements}
This work was supported in part by National Natural Science Foundation of China (grant numbers 62131015, 62250710165, U23A20295), the STI 2030-Major Projects (No. 2022ZD0209000), Shanghai Municipal Central Guided Local Science and Technology Development Fund (grant number YDZX20233100001001), Science and Technology Commission of Shanghai Municipality (STCSM) (grant number 21010502600), The Key R\&D Program of Guangdong Province, China (grant numbers 2023B0303040001, 2021B0101420006), and Science and Technology special fund of Hainan Province (grant number KJRC2023B06).

\section*{Author contributions statement}

Y.W. conducted study design, algorithm implementation, experimental setup, data collection, data processing, and manuscript writing.
H.X. participated in algorithm implementation and manuscript writing.
K.S. participated in experimental setup and manuscript revision and editing.
L.D. participated in manuscript review and editing.
S.B. evaluated experimental results and provided clinical feedback on the study.
Z.D. evaluated experimental results and provided clinical feedback on the study.
J.L. helped collect and process data used for image synthesis model training and text-encoder pre-training.
Q.W. supervised the research.
Q.L. supervised the research, conducted funding acquisition and provided manuscript revision.
D.S. supervised the entire research by providing ideas and guiding detailed implementation and experiments, conducted funding acquisition, and provided the detailed clue for writing as well as manuscript revision.  


\begin{sidewaystable}[]
\normalsize
\caption{Description and characteristics of our collected brain MRI database including 31,407 scans from 13 data centers.}
\label{dataset}
{%
\begin{tabular}{cccccccccclcccc}
\hline
                                                        & \multicolumn{9}{c}{Internal datasets}                                                                                                                                                                                                                                                                                                                                                                                                                                                                                                                                               &  & \multicolumn{4}{c}{External datasets}                                                                                                                                                                  \\ \cline{2-10} \cline{12-15} 
Datasets                                                & OASIS3                                                                    & HCPY                                                    & IXI                                                    & HCPD                                                    & BCP                                                     & AIBL                                                           & HCPA                                                    & CBMFM                                                     & \begin{tabular}[c]{@{}c@{}}BraTS\\ 2021\end{tabular}                      &  & ADNI-2                                                  & \begin{tabular}[c]{@{}c@{}}UK\\ Biobank\end{tabular} & ABCD                                                    & Ren Ji                                               \\
Modality                                                & \begin{tabular}[c]{@{}c@{}}T1w\\ T2w\\ FLAIR\\ SWI\\ T2 Star\end{tabular} & \begin{tabular}[c]{@{}c@{}}T1w\\ T2w\end{tabular}       & \begin{tabular}[c]{@{}c@{}}T1w\\ T2w\\ PD\end{tabular} & \begin{tabular}[c]{@{}c@{}}T1w\\ T2w\end{tabular}       & \begin{tabular}[c]{@{}c@{}}T1w\\ T2w\end{tabular}       & \begin{tabular}[c]{@{}c@{}}T1w\\ T2w\\ FLAIR\\ PD\end{tabular} & \begin{tabular}[c]{@{}c@{}}T1w\\ T2w\end{tabular}       & \begin{tabular}[c]{@{}c@{}}T1w\\ T2w\\ FLAIR\end{tabular} & \begin{tabular}[c]{@{}c@{}}T1w\\ T2w\\ FLAIR\\ T1CE\end{tabular}          &  & \begin{tabular}[c]{@{}c@{}}T1w\\ FLAIR\end{tabular}     & \begin{tabular}[c]{@{}c@{}}T1w\\ FLAIR\end{tabular}  & \begin{tabular}[c]{@{}c@{}}T1w\\ T2w\end{tabular}       & \begin{tabular}[c]{@{}c@{}}T1w\\ FLAIR\end{tabular} \\
\begin{tabular}[c]{@{}c@{}}Image\\ Numbers\end{tabular} & 5140                                                                      & 2220                                                    & 1476                                                   & 1304                                                    & 340                                                     & 3383                                                           & 1450                                                    & 2550                                                      & 3788                                                                      &  & 1000                                                    & 2000                                                 & 5728                                                    & 1028                                                \\
\begin{tabular}[c]{@{}c@{}}Pair\\ Numbers\end{tabular}  & 800                                                                       & 1110                                                    & 492                                                    & 652                                                     & 170                                                     & 208                                                            & 725                                                     & 839                                                       & 947                                                                       &  & 500                                                     & 1000                                                 & 2864                                                    & 514                                                 \\
Age                                                     & 42-95                                                                     & 22-35                                                   & 20-86                                                  & 5-21                                                    & 2-5                                                     & 60-89                                                          & 35-100+                                                 & 18-69                                                     & 18-87                                                                     &  & 49-97                                                   & 44-82                                                & 9-19                                                    & 35-84                                               \\
Regions                                                 & \begin{tabular}[c]{@{}c@{}}North\\ America\end{tabular}                   & \begin{tabular}[c]{@{}c@{}}North\\ America\end{tabular} & Europe                                                 & \begin{tabular}[c]{@{}c@{}}North\\ America\end{tabular} & \begin{tabular}[c]{@{}c@{}}North\\ America\end{tabular} & Oceania                                                        & \begin{tabular}[c]{@{}c@{}}North\\ America\end{tabular} & Asia                                                      & \begin{tabular}[c]{@{}c@{}}North\\ America,\\ Europe,\\ Asia\end{tabular} &  & \begin{tabular}[c]{@{}c@{}}North\\ America\end{tabular} & Europe                                               & \begin{tabular}[c]{@{}c@{}}North\\ America\end{tabular} & Asia                                                \\ \hline
\end{tabular}
}
\end{sidewaystable}

\begin{table}[]
\caption{Performance of zero-shot synthesis by our TUMSyn, the T1w-specific synthesis model SynthSR, and model variants using one-hot text encoding and BiomedCLIP text encoding on four external validation datasets across eight synthesis tasks in terms of PSNR and SSIM. The non-zero-shot synthesis by SC-GAN (particularly trained on these datasets) is integrated here as benchmark.}
\label{tab:table1}
\resizebox{\columnwidth}{!}{
\begin{tabular}{ccccccc}
\hline
                            &                                          & SC-GAN        & SynthSR       & One-hot       & BiomedCLIP    & TUMSyn     \\ \cline{3-7} 
 {ABCD}       &  {T1w-\textgreater{}T2w}   & \textbf{29.17 ± 1.12}  & --            & 23.01 ± 1.52  & 25.57 ± 1.03  & 28.79 ± 1.25  \\
                            &                                          & \textbf{0.969 ± 0.009} & --            & 0.898 ± 0.023 & 0.939 ± 0.018 & 0.967 ± 0.009 \\
                            &  {T2w-\textgreater{}T1w}   & \textbf{26.80 ± 0.97}  & 19.44 ± 1.07  & 24.81 ± 0.89  & 25.12 ± 0.83  & 26.47 ± 1.16  \\
                            &                                          & \textbf{0.951 ± 0.016} & 0.806 ± 0.018 & 0.936 ± 0.009 & 0.941 ± 0.008 & 0.944 ± 0.016 \\
 {ADNI-2}     &  {T1w-\textgreater{}FLAIR} & 23.50 ± 0.73  & --            & 24.85 ± 0.91  & 25.19 ± 0.88  & \textbf{26.63 ± 0.95}  \\
                            &                                          & 0.929 ± 0.007 & --            & 0.944 ± 0.008 & 0.952 ± 0.007 & \textbf{0.958 ± 0.007} \\
                            &  {FLAIR-\textgreater{}T1w} & 24.96 ± 0.80  & 23.29 ± 1.31  & 26.02 ± 0.83  & 26.18 ± 0.69  & \textbf{26.85 ± 0.74}  \\
                            &                                          & 0.944 ± 0.009 & 0.928 ± 0.026 & 0.953 ± 0.007 & 0.957 ± 0.007 & \textbf{0.960 ± 0.006} \\
 {UK Biobank} &  {T1w-\textgreater{}FLAIR} & \textbf{31.02 ± 0.79}  & --            & 26.39 ± 0.53  & 28.76 ± 0.76  & 30.09 ± 0.61  \\
                            &                                          & \textbf{0.976 ± 0.004} & --            & 0.949 ± 0.012 & 0.962 ± 0.009 & 0.972 ± 0.004 \\
                            &  {FLAIR-\textgreater{}T1w} & \textbf{30.34 ± 0.85}  & 18.71 ± 0.69  & 25.65 ± 0.54  & 27.05 ± 1.12  & 29.81 ± 0.82  \\
                            &                                          & \textbf{0.973 ± 0.005} & 0.832 ± 0.017 & 0.945 ± 0.007 & 0.967 ± 0.018 & 0.970 ± 0.007 \\
 {Ren Ji}      &  {T1w-\textgreater{}FLAIR} & \textbf{26.24 ± 1.61}  & --            & 23.05 ± 1.12  & 24.02 ± 1.33  & 26.00 ± 1.53  \\
                            &                                          & \textbf{0.956 ± 0.016} & --            & 0.897 ± 0.018 & 0.920 ± 0.017 & 0.947 ± 0.017 \\
                            &  {FLAIR-\textgreater{}T1w} & \textbf{28.28 ± 1.53}  & 18.87 ± 1.26  & 23.42 ± 0.50  & 25.30 ± 0.96  & 28.02 ± 1.45  \\
                            &                                          & \textbf{0.964 ± 0.018} & 0.847 ± 0.036 & 0.939 ± 0.012 & 0.945 ± 0.013 & 0.960 ± 0.015 \\ \hline
\end{tabular}
}
\end{table}
\end{document}